\documentclass[final,5p,times]{elsarticle}
\usepackage{graphicx}
\usepackage{txfonts}

\usepackage{inputenc}
\usepackage{hyperref}
\usepackage{amssymb}
\usepackage{xcolor}
\usepackage{soul}

\journal{Journal of Computational Physics}

\newcommand{\firstReview}[1]{ {\color{black}{#1}} }
\newcommand{\secondReview}[1]{ {\color{black}{#1}} }

\begin{document} 

	\begin{frontmatter}

   		\title{LightAMR format standard and lossless compression algorithms for adaptive mesh refinement grids: RAMSES use case}

		\author[AIM,IRFU]{L.Strafella}
		\ead{loic.strafella@cea.fr}
		
		\author[IRFU]{D.Chapon}
		\ead{damien.chapon@cea.fr}
		
		\address[AIM]{AIM, CEA, CNRS, Universit\'e Paris-Saclay, 91191 Gif-sur-Yvette, France}
		\address[IRFU]{IRFU, CEA, Universit\'e Paris-Saclay, 91191 Gif-sur-Yvette, France}		
                         
		\begin{abstract}
			The evolution of parallel I/O library as well as new concepts such as ’in transit’ and ’in situ’ visualization and analysis have been identified as key technologies to circumvent I/O bottleneck in pre-exascale applications. Nevertheless, data structure and data format can also be improved for both reducing I/O volume and improving data interoperability between data producer and data consumer. In this paper, we propose a very lightweight and purpose-specific post-processing data model for AMR meshes, called \textit{lightAMR}. Based on this data model, we introduce a tree pruning algorithm that removes data redundancy from a fully threaded AMR octree. In addition, we present two lossless compression algorithms, one for the AMR grid structure description and one for AMR double/single precision physical quantity scalar fields. Then we present performance benchmarks on RAMSES simulation datasets of this new lightAMR data model and the pruning and compression algorithms. We show that our pruning algorithm can reduce the total number of cells from RAMSES AMR datasets by 10-40\% without loss of information. Finally, we show that the RAMSES AMR grid structure can be compacted by $\sim 3$ orders of magnitude and the float scalar fields can be compressed by a factor $\sim 1.2$ for double precision and $\sim 1.3-1.5$ in single precision with a compression speed of $\sim 1$ GB/s. 
			
		\end{abstract}

   \end{frontmatter}

\section{Introduction}

RAMSES is a massively parallel hydrodynamical code for self-gravitating magnetized flows widely used in the computational astrophysics community \citep{Teyssier2002}, \firstReview{to solve the evolution of dark matter, stellar populations, and gas via gravity, hydrodynamics, radiative transfer, and non-equilibrium radiative cooling/heating. For hydrodynamics, it uses the HLLC Riemann solver \citep{Toro1994} and the MinMod slope limiter to construct gas variables at cell interfaces from their cell-centered values. The dynamics of collision-less dark matter and star particles are evolved with a multi-grid particle-mesh solver and cloud-in-cell interpolation \citep{Guillet2011}.}

It implements an adaptative mesh refinement (AMR) technique to optimize the memory/precision ratio in numerical simulations, \cite{Berger89}, \cite{Berger84}. Its data structure is a cell-based fully threaded octree \citep{Khokhlov_1998}, \firstReview{which means that each subdomain describes a full octree of the entire simulation domain up to its topmost root grid which has the size of the whole simulation domain. In this distributed AMR tree approach each subdomain defines its own local grid refinement.} In most use cases, the domain decomposition is based on a 1D Hilbert curve to split the calculation between MPI processes. \firstReview{It was first designed for cosmological and galaxy numerical simulation, and was used for large simulation \citep{Teyssier2009}, \citep{Ocvirk2020}, \citep{Chabanier2020}.} However, RAMSES faces I/O scalability bottlenecks when used over a few thousand of MPI processes, partly due to its "multiple file per process" I/O strategy. A typical cosmological simulation using 10,000 MPI processes will produce at least 40,000 files (and filesystem inodes) per output with small file size from dozens of megabytes to a few hundred megabytes, which is far from optimal on modern Lustre filesystems and generally rapidly reach the user's limits on supercomputer. \firstReview{What is more, since RAMSES only implements one output data format, for both checkpointing and post-processing purposes, it is not optimized for both cases. Therefore, a simulation like Extreme-Horizon \citep{Chabanier2020} produced 3.6 terabytes of data with more than 50,000 files per snapshot (AMR files, hydrodynamical quantity files, particles files, etc). Considering that several snapshots are required for scientific analysis that kind of simulation becomes rapidly challenging in terms of I/O, data management and post-processing.}\\

In a previous work, the integration of the the Hercule library \citep{bressand} for parallel I/O and data management in RAMSES was the first step to improve the I/O performance, scalability and data management \citep{Strafella2020_Astronum}. \firstReview{In terms of data management, the integration of a parallel I/O library was also the occasion to add a post-processing specific data-flow to RAMSES. This was an important step which permitted to optimize significantly the checkpoint dump since the checkpoint's raw data will no longer be used by post-processing tools for simulation data analysis. With the original version of RAMSES, sometimes the size of the simulation is calculated not based on CPU performance, memory consumption, or pure scalability of the code but simply based on the user's disk storage available capacity. We expect this problem to occur more and more often in the coming years. The usually adopted solution is either to reduce the size of the simulation, the precision of the simulation by tuning AMR parameters or reducing the number of produced snapshots. The question was then is there a standard and compact data model that can be used to describe the AMR data from RAMSES new post-processing data-flow ?}


\firstReview{Hercule library's post-processing database strategy is to propose an XML dictionary containing the description of standardized data model for different kind of meshing strategies and objects. Therefore, once a standardized data model is described in the dictionary, every tool using Hercule will be able to "understand" Hercule post-processing data. As a reminder, we show in table \ref{tab:ramses-io} a small recap of I/O data-flows now available in RAMSES.} 


%
\begin{table}[!h]
	\begin{tabular}{l|ccc}
		\hline
		I/O library & Posix & Hercule & Hercule \\
		\hline
		Purpose & C/R+ P-P & C/R & P-P\\
		Data Structure & Specific & Free & Standardized \\
		Data Volume & Large & Large & Light \\
		Compression & N/A & N/A & Efficient\\
		\hline
	\end{tabular}
	\caption{Difference between various RAMSES I/O formats : the legacy binary posix and the new Hercule database HProt, HDep. C/R: Checkpoint/Restart, P-P: Post-processing.}
	\label{tab:ramses-io}
\end{table}

\firstReview{Until a few years ago, AMR data did lack of a standardized format for post-processing and visualization purposes partly due to the different flavor of AMR strategies: block-based / patched-based \cite{Gamer-code}, \cite{Enzo-code}, \cite{Flash-code}, cell-based (RAMSES), etc. Thus AMR data still required some data transformation step (e.\ g.\ down-casting to unstructured grids, resampling to fixed resolution cartesian grids) to be manipulated by general purpose post-processing or visualization tools (e.\ g.\ Paraview \footnote{\href{https://www.paraview.org/}{Paraview.org}}, VisIt \footnote{\href{https://wci.llnl.gov/simulation/computer-codes/visit}{VisIt home page}}).} An AMR mesh described using unstructured grids can easily lead to memory issues and produce high volume data files, \firstReview{thus each AMR code uses a custom data model to output the data. In the case of RAMSES, the AMR internal 32 bits integer linked-lists are dumped level by level.} Code-specific data-processing libraries were then implemented as solutions to read the RAMSES datafile format and handle its AMR grid structures (e.\ g.\ PyMSES\footnote{\href{http://irfu.cea.fr/Projets/PYMSES}{http://irfu.cea.fr/Projets/PYMSES}}, Osiris \footnote{\href{https://www.nbi.dk/~nvaytet/osiris/osiris.html}{https://www.nbi.dk/~nvaytet/osiris/osiris.html}})
but these libraries needed to be updated to follow the RAMSES data format evolutions \citep{Chapon2013}. \firstReview{A standardized and self-consistent data model aims at reducing the cost of post-processing tools maintenance and encourage sharing development efforts.} \\


In this work, we introduce a new lightweight AMR data model that we applied on RAMSES AMR data. By using this data model, we show that data volume of RAMSES outputs  can be significantly reduced, \firstReview{without loss of information}. In addition, we present two lossless data compression algorithms that can be built upon this new data model standard to further reduce both the size of the AMR grid refinement description and the size of associated physical quantity scalar field data (float 32/64 bits). In the results section, we present \firstReview{ lightAMR data volume reduction as well as tree pruning and} compression benchmarks on published RAMSES simulation datasets. 

\section{Scientific analysis-specific AMR light tree data model}
\label{sec:hdepformat}

\subsection{Motivation}

The RAMSES checkpoint/restart data format was designed, as in many other simulation codes used in astrophysics, to hold all the information necessary for the code to restart a simulation run and resume the dynamical evolution of the numerical model. Since our goal is to design a new output data format for the specific purpose of scientific analysis and post-processing, we are free to select from all the information available in memory at runtime what is strictly necessary to answer one's post-processing need.\\

To fit one's post-processing needs, a user can choose to export a subset of physical quantities (e.\ g.\ store only the density and thermal pressure scalar fields defined on the AMR grid, and discard the gravitational potential/acceleration AMR fields, the velocity AMR vector field and the Lagrangian particle information), hence reducing the overall dataset volume on disk. In RAMSES, this user-defined data field subset selection can be configured at runtime using the input parameter namelist, without the need to recompile the code. What is more, for some data post-processing use cases, down-casting these quantities from double to single precision could even prove sufficient, reducing further the data volume by a factor of 2. Finally, since this new data format is not meant to be read by RAMSES itself upon restart but only by data post-processing tools, we can decide to adopt a new standardized AMR grid data structure and a custom encoding to be able to answer any type of post-processing software requirement in an optimized and standardized way.

\subsection{Implementation}
\begin{figure}[!h]
   \centering
   \includegraphics[width=\columnwidth]{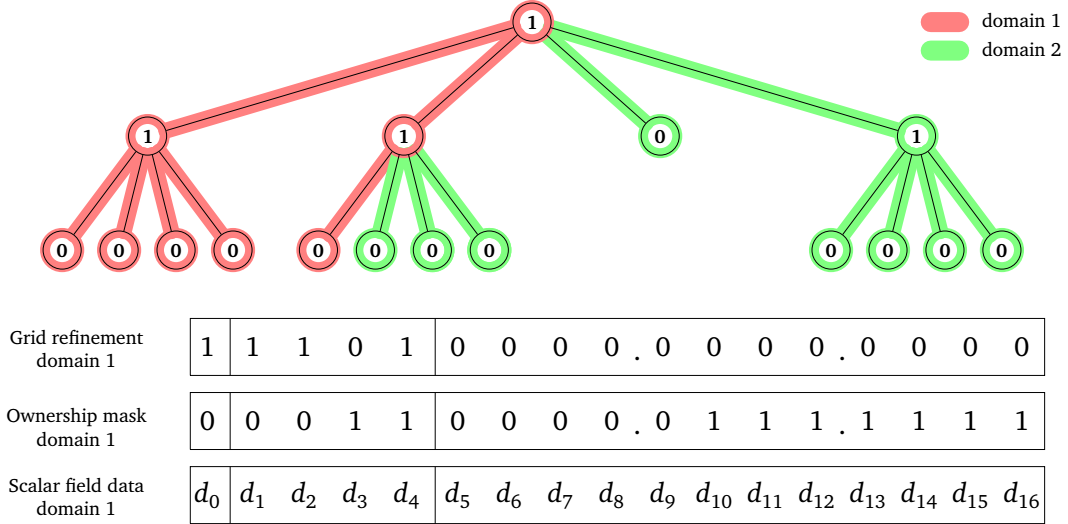}
   \caption{Description of a 2D AMR grid (with a refinement factor of 2) dispatched on two MPI processes according to the lightAMR standardized data model (above). Grid refinement, ownership mask and scalar data field arrays for the domain \#1 (below).}
   \label{fig:amr_grid}
\end{figure}

The Hercule I/O library HDep standardized data model for AMR trees aims to offer a light and self-consistent way to describe an AMR grid. In this work, we adopt a similar approach as the one proposed in the HDep format, the details of our approach is described here. We further reference this standardized data model for AMR grids as \textit{lightAMR}. Since that kind of mesh can be expressed in the form of a tree, it uses a boolean description of the AMR tree from top to bottom, level by level, as shown on figure \ref{fig:amr_grid}. With this breadth first traversal description of an AMR tree, two main arrays are used: one for the grid refinement description and one for the cell ownership mask.
The first array describes the refinement state of a cell: 1 for a refined cell and 0 for a leaf (unrefined) cell. All child cells from a refined one must be described, as shown in \ref{fig:amr_grid}. This implies that the number of cells at a level $l$ must be equal to the number of refined cells at level $l-1$ times the refinement factor to the power of the dimension. In our 2D example, the refinement factor being 2 by dimension, the level 2 must contain $3\times2^2$ boolean values since there are 3 refined cells at level 1. The value in the ownership mask array describes for each cell if it does not belong to the current domain (according to the domain decomposition policy): 0 if it belongs to the current domain, otherwise 1.

In order to be self-consistent, a few other (lightweight) meta-information such as the size of the simulation box, the physical quantity units, physical model details  or used numerical schemes need to be stored, using an explicit semantic (string key-value pairs). In addition, some useful meta-information like the number of cells per level, the total number of grids, the number of levels, if somewhat redundant with the AMR grid description, can be conveniently stored to enhance the post-processing tools performance. AMR-bound physical quantities are described in scalar field data arrays that follow the same ordering as the grid refinement array, hence providing a straightforward mapping (which facilitates the LOD approach). Since RAMSES describes the AMR tree from the root node which is the simulation box itself on each domain and stores scalar field values even for refined (coarse) cells, the construction of these arrays is straightforward when converting in the lightAMR format. Of course, this lightAMR tree data format can be easily extended to be compatible with other numerical codes that do not share RAMSES fully-threaded tree data structure, refined cell (coarse) scalar field values only requiring to be computed on-the-fly upon data export.\\

Collision-less particles are treated in a Lagrangian way within RAMSES and are linked to the AMR grid. The particles and their associated properties are simply added to our new data output format in the form of one dimensional arrays. Particle positions for example are in the form $x_{1},y_{1},z_{1},...,x_{n},y_{n},z_{n}$. We chose not to associate particles to their linked AMR cells and consequently not overload the data format. The particle-cell binding is left to the post-processing tools to be performed on-the-fly if needed.\\

For simulation codes that do not use the fully-threaded approach and naturally describe collections of trees at a defined minimum level $l_0$ on simulation domains, the LightAMR data model can be trivially extended. One just needs to provide an additional AMR root description array to detail the I-J-K logical position indices of the root coarse cells of those trees in a structured grid at level $l_0$. While providing an array containing the logical position in $(i_1,j_1,k_1, i_2,j_2, k_2, ...)$ format would be the natural approach, one may choose to provide an array of Morton indices of logical position at the level $l_0$. Indeed, using the Morton curve, 3D/2D logical coordinates can be mapped on 1D Morton curve by bits interleaving, hence reducing the size of the array by a factor 3 (resp. 2) in 3D (resp. 2D) but with a limitation on the maximum value of $l_0$ 32 bits Morton indices: $l_{0,\textrm{max}} = 10$ in 3D ($l_{0,\textrm{max}} = 16$ in 2D). Within this extension, compared to their fully-threaded tree version, the grid refinement, ownership mask and scalar field data arrays should be trimmed up to level $l_0$ as a consequence.



\section{Tree pruning to remove redundancy}

Due to the fully threaded octree approach in RAMSES, some AMR cells may be described several times in the trees of different domains, leading to data redundancy. Even if those multiple AMR cell descriptions are required by a numerical solver like the Poisson equation multi-grid solver \citep{Guillet2011}, describing those cells only once is sufficient for post-processing and visualization purposes. Therefore, a tree pruning algorithm can be implemented to reduce the memory footprint of our file format.

\begin{figure}[!h]
   \centering   
   \includegraphics[width=\columnwidth]{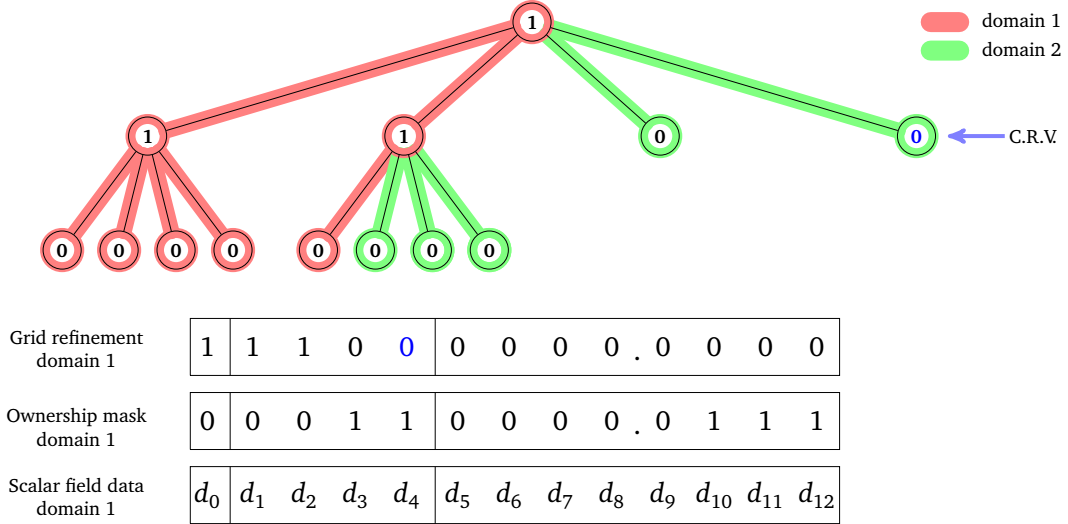}
   \caption{Applying a tree pruning algorithm on the AMR tree of figure \ref{fig:amr_grid} for domain \#1 (red), and Change of Refinement Value (C.R.V) for an unnecessary refined cell for the domain 1 . Notice the change of value in the grid refinement array in
   blue, as well as the propagation of this change on the ownership mask and scalar field data arrays (truncation of the arrays after $d_{12}$).}
   \label{fig:pruning}
\end{figure}

Once the tree is built according to lightAMR format described above, most of the data redundancy in the AMR dataset can be removed. We developed a tree pruning algorithm, propagating from the bottom of the tree to its top, in order to remove the redundant cells. The unnecessary cells are removed by changing their refinement value (C.R.V on figure \ref{fig:pruning}) and removing their described children cells on the next level. The C.R.V. is done only for a refined cell if all its children belong to another domain (mask = 1), as shown on figure \ref{fig:pruning}.

Then those changes are propagated to the associated arrays: the ownership mask and scalar field data arrays. Therefore, removing those refined cells and its children is the first step to reduce the total volume of data before applying any compression algorithm. Depending on the simulation, this step can prune a significant part of any RAMSES simulation dataset, see results on astrophysical simulation datasets in section \ref{subsec:results_pruning}. \firstReview{Prior to the tree pruning step, a temporary mask array needs to be adapted from the original ownership mask array to properly compute the C.R.V. This slight adaptation insures that any refined cell is not flagged as masked if at least one of its children is not masked. This temporary mask is only used during the tree pruning step}.\\



At the end of this step, the AMR grid is properly described with one or several trees and most of the data redundancy was removed. However, for a fully-threaded octree the redundancy cannot be entirely removed unless the tree is split at the coarse level of refinement (a.k.a. $levelmin$ in RAMSES) and domain boundaries set between sibling cells at that coarse level (and not deeper in the tree). The next stage in the lightAMR formatting is to apply data compression on both the grid refinement/ownership mask arrays and the simple/double precision associated scalar field data arrays.

\section{AMR data compression}

To overcome the I/O bottleneck, in addition to the integration of parallel I/O  \citep{Strafella2020_Astronum} and a data reduction strategy, data compression can reduce the volume of data to transfer or store on disk/object store and further improve the overall I/O performance. Combining those different strategies can have a significant impact on the I/O times, data volume storage and data transfer rates, whether for in-transit or in-situ data post-processing. In this section, we present two lossless compression schemes specific to octree AMR meshes that we developed for the lightAMR data model; one for the grid description boolean valued arrays and one for the scalar field float (32/64 bits) data arrays.

\subsection{LightAMR grid boolean arrays lossless compression}
\label{sec:compression_amr_struct}

Once the lightAMR structure has been described with boolean valued arrays, we developed a deterministic, lossless and performant algorithm to compress the information contained in the two main arrays (grid refinement and ownership mask arrays) based on \secondReview{a custom} run-length encoding \firstReview{which is well known compression technique \citep{Robinson1967}}. The first step of the algorithm is to consider the boolean array as an ordered list of packs of identical bit values, either successive zeroes or successive ones, and count the lengths of these continuous packs. The second step is to digitize these ordered pack sizes with a special encoding and store the result in a single byte integer array. Out of the 256 integer values available, our encoding only need values between 1 and 62. To encode pack sizes, we use a base-52 decomposition and we also record the number of digits (only if greater than 1) in that decomposition. The uint8 values ranging from 2 to 7 are used to store the number of bytes used for the encoding of a each pack size, and the 52 uint8 values between 11 and 62 (values 9 and 10 are used for specific purposes, see next paragraph) to store the digits of the base-52 decomposition (with an additional shift of $11$). For example, to encode a pack size of 2904 which decompose in $1\cdot52^2 + 3\cdot52^1 + 44\cdot52^0$, the number of digit (3) is stored, followed by the digits $[1,3,44]$, encoded as $[12,14,55]$ ($[1+11, 3+11, 44+11]$), resulting in the following 4 bytes $[3,12,14,55]$ instead of the 262 bytes required to store 2904 zeroes or ones.

\begin{figure}[!h]
	\centering
    \includegraphics[width=\columnwidth]{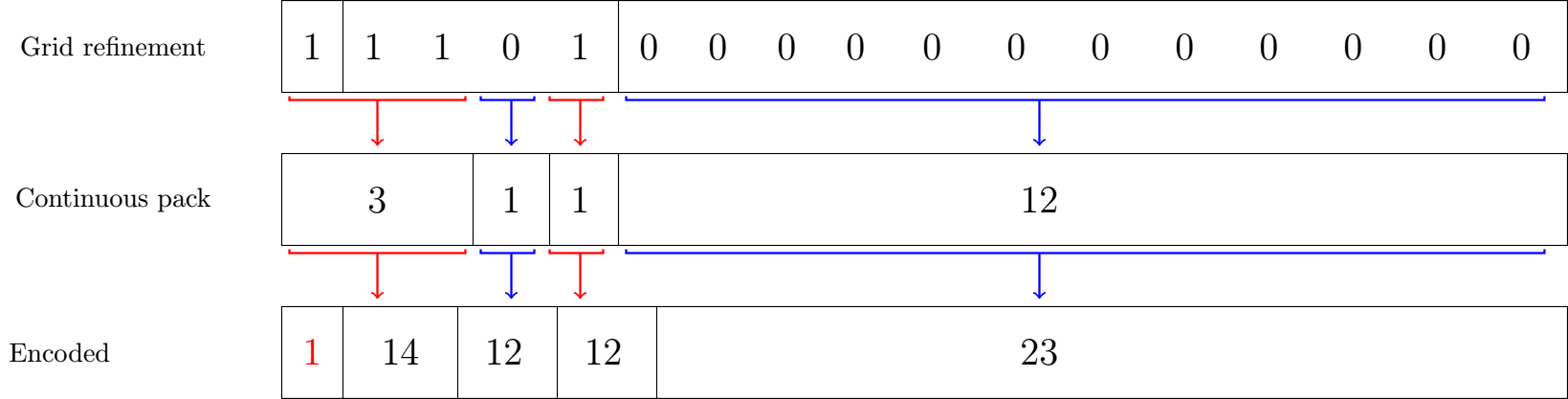}
	\caption{AMR grid refinement boolean array encoding in base 52 after counting the sizes of continuous packs of successive 0 and 1. The bit value of the first pack is encoded in red.}
	\label{fig:compression_amr1}
 \end{figure}

With that encoding, the greatest pack size that can be encoded is $52^7-1 \sim 10^{12}$ continuous 0 or 1, which is more than enough, the maximum number of cells per domain in a typical RAMSES simulation run never exceeding a $ 10^{7}$ due to memory limitations. Since the encoded pack sizes describe identical bit values, alternatively contiguous zeroes and contiguous ones, only the first value of the array needs to define the bit value of the first pack (see the red value in figure \ref{fig:compression_amr1}), and all subsequent bit values of the following packs can be determined by the decompression algorithm.\\

\begin{figure}[!h]
   \centering
   \includegraphics[width=\columnwidth]{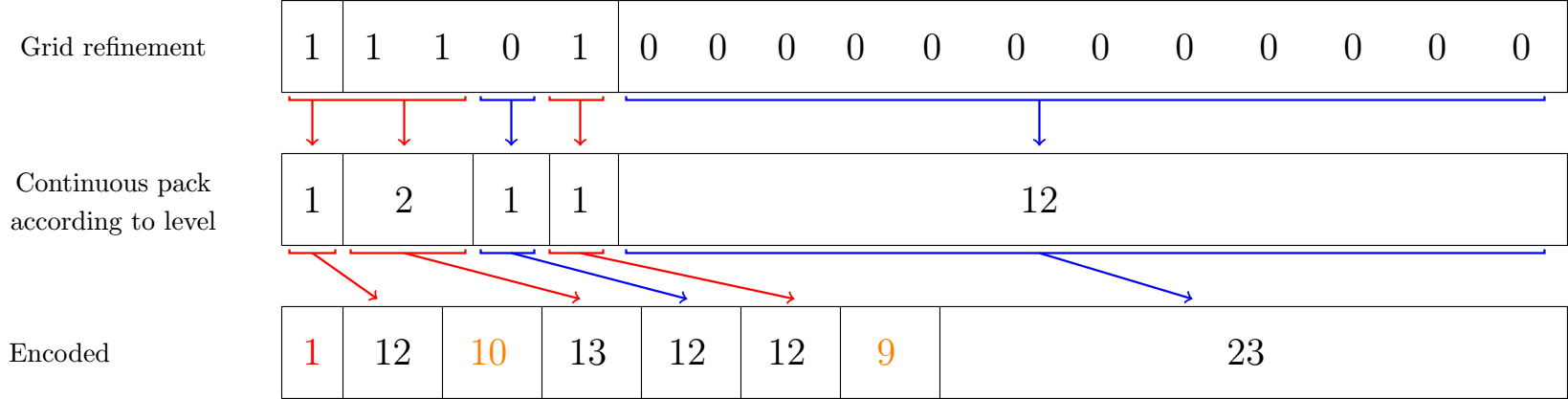}
   \caption{Encoding of AMR description array by counting the size of continuous pack of 0 and 1 according to the number of cells by level, and adding markers 10 or 9 according to a potential bit value flip.}
   \label{fig:compression_amr2}
\end{figure}

To give the possibility to data-processing tools to improve their memory footprint, we included special values in this encoding to mark the boundaries of AMR levels so that data-processing tools can perform on-the-fly decompression on a \textit{level-by-level} basis for level-of-details approach. We used the 9 and 10 uint8 values as level boundary markers, two different values being required to mark level boundaries in case a bit flip occur (or not) at the end of an AMR level (see orange values in figure \ref{fig:compression_amr2}). In addition, we developed a special algorithm which is able to directly process the compressed stream to get the value at a given offset in the uncompressed stream without the need of uncompressing the entire AMR compressed array.

We further reference this LightAMR grid boolean arrays lossless compression as base 52 continuous pack size encoding (CPS52). \secondReview{While this CPS52 encoding may seem peculiar compared to the majority of other RLE schemes from the literature, it yields very good performance. In the process of LightAMR data format standardisation though, this custom RLE would require to be updated to match more closely standard encoding schemes, prior to the LightAMR format standard release.} Once the AMR grid description arrays are compressed, the next step in order to reduce the data volume is to compress the float (simple/double precision) scalar field data arrays.

\subsection{LightAMR scalar field float data lossless compression}
\label{sec:compression_amr_data}

The development of our custom compression algorithm was motivated by the poor compression rates obtained with the deflate algorithm from the zlib \footnote{https://zlib.net/}, about 3 to 5\% with a compression speed about 50 MB/s. Even if higher compression rates could be achieved using the LZMA algorithm, it would require far more memory for a lower compression speed, 
such memory-hungry approach would not be viable in memory-bound massively parallel simulations as are many RAMSES runs. In addition, compression of float valued arrays using entropic encoding without prior knowledge of the data generally leads to poor performance due to its intrinsic homogeneous byte distribution unlike for example compressing a text where the distribution of byte (letters) is not homogeneous and therefore entropic compression is highly efficient.\\

In this section we present an adaptation of delta compression algorithm for lossless float data \citep{Lindstrom2006}, \citep{Burtscher2009} that we specifically designed for lightAMR octree structures. Delta compression is a commonly used compression method based on prediction functions. Indeed, this method is based on the choice of a mathematical predictor function and encode directly the prediction error. The compression process itself is performed by removing the leading zeroes in the encoding of the prediction error \citep{Lindstrom2006}.

With this kind of approach, high compression ratios are reached when the predictor function is quite accurate, which is not easy to achieve when dealing with an AMR octree structure. Thus, we extended this delta compression algorithm to AMR grids by taking advantage of the fully threaded approach of the AMR strategy in RAMSES. \\

\begin{figure}[!h]
	\centering
    \includegraphics[width=\columnwidth]{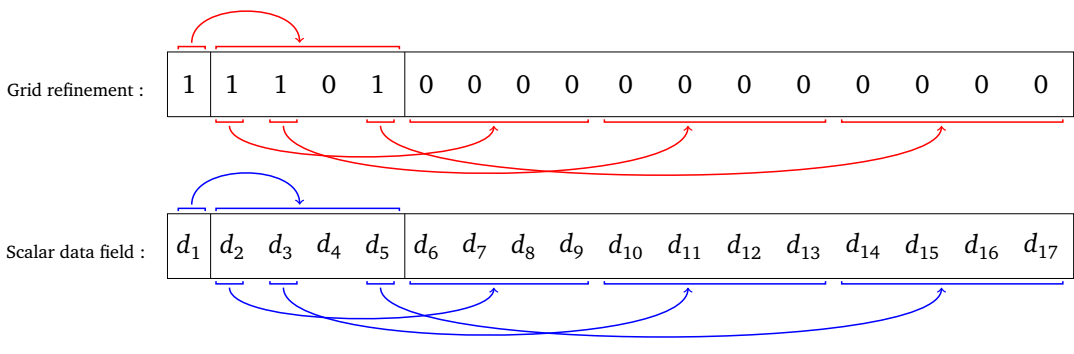}
	\caption{Parent-Child predictor : the parent cell value is used to predict the child cell value in order to achieve delta compression.}
	\label{fig:fsp_compression}
 \end{figure}

The only information our compression algorithm needs is the grid refinement array and a physical quantity field to compress (simple or double precision float). We propose to use an AMR-specific mathematical function to predict intensive variable scalar field values called the \firstReview{\textit{Parent-Child predictor} (PCP)} function. Indeed, the value of the parent cell is used as the predicted value for its child cells and so on from the top of the AMR tree down to the bottom, as represented on figure \ref{fig:fsp_compression}. The compression is then made octant by octant (if 3 dimensions are used). There are two mains steps in the data compression process: first, the delta compression itself with the leading zeroes removal and second, the encoding of the compressed scalar field \firstReview{on a bit-field}.\\

\begin{figure}[!h]
	\centering
	\includegraphics[width=\columnwidth]{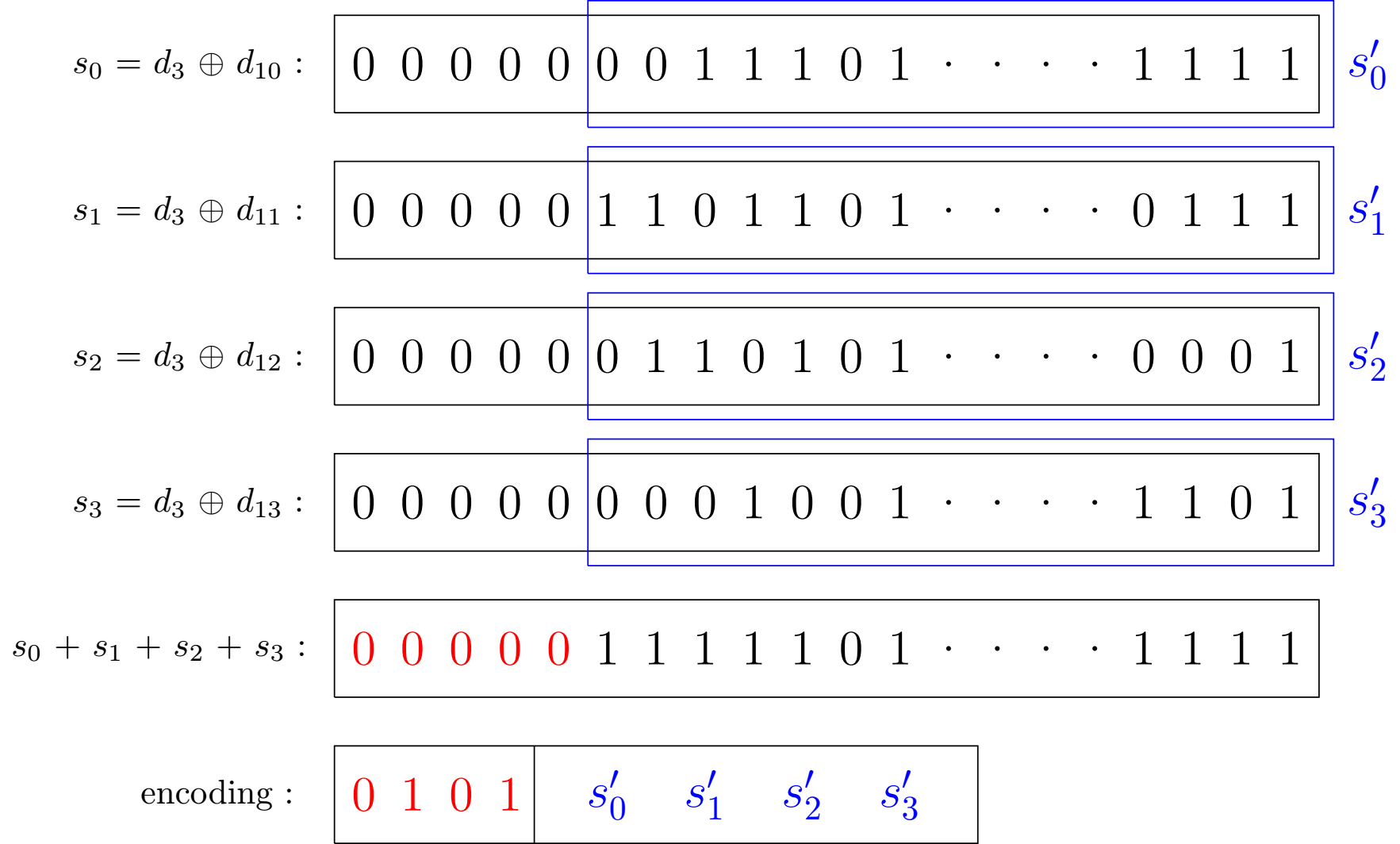}
	\caption{XOR operation (sign \firstReview{$\oplus$}) between the \firstReview{parent} cell value $d3$ and its children values $d_{10},d_{11},d_{12},d_{13}$, and the \firstReview{OR} (sign \firstReview{$+$}) operation between $s_{0},s_{1},s_{2},s_{3}$ in order to compute the number of removable leading zeroes, in red. Residues of $s_{i}$ values are noted $s^{'}_{i}$.}
	\label{fig:fsp_encoding}
 \end{figure}

As for the delta compression stage, the single (or double) precision parent float value of the physical field is mapped on unsigned integer of the same size (8 bytes unsigned integer for a double precision float and 4 bytes for a single precision) as well as the values of the children cells. Then we use an XOR operator to compute the differences between the mapped parent value and all the children ones, leading to a pack of prediction errors ($s_{1}, s_{2}, s_{3},s_{4}$ in the 2D example shown on figure \ref{fig:fsp_encoding}). The next step is to compute \firstReview{an OR} operation between all the $s_{i}$ values in order to determine how many leading zeroes could be removed on each $s_{i}$ value at most, which corresponds to the stage $s_0 + s_1 + s_2 + s_3$ in figure \ref{fig:fsp_encoding}. In the example, we can remove a maximum of 5 leading zeroes on each $s_{i}$ value without losing information. 

\begin{figure}[!h]
	\centering
    \includegraphics[width=\columnwidth]{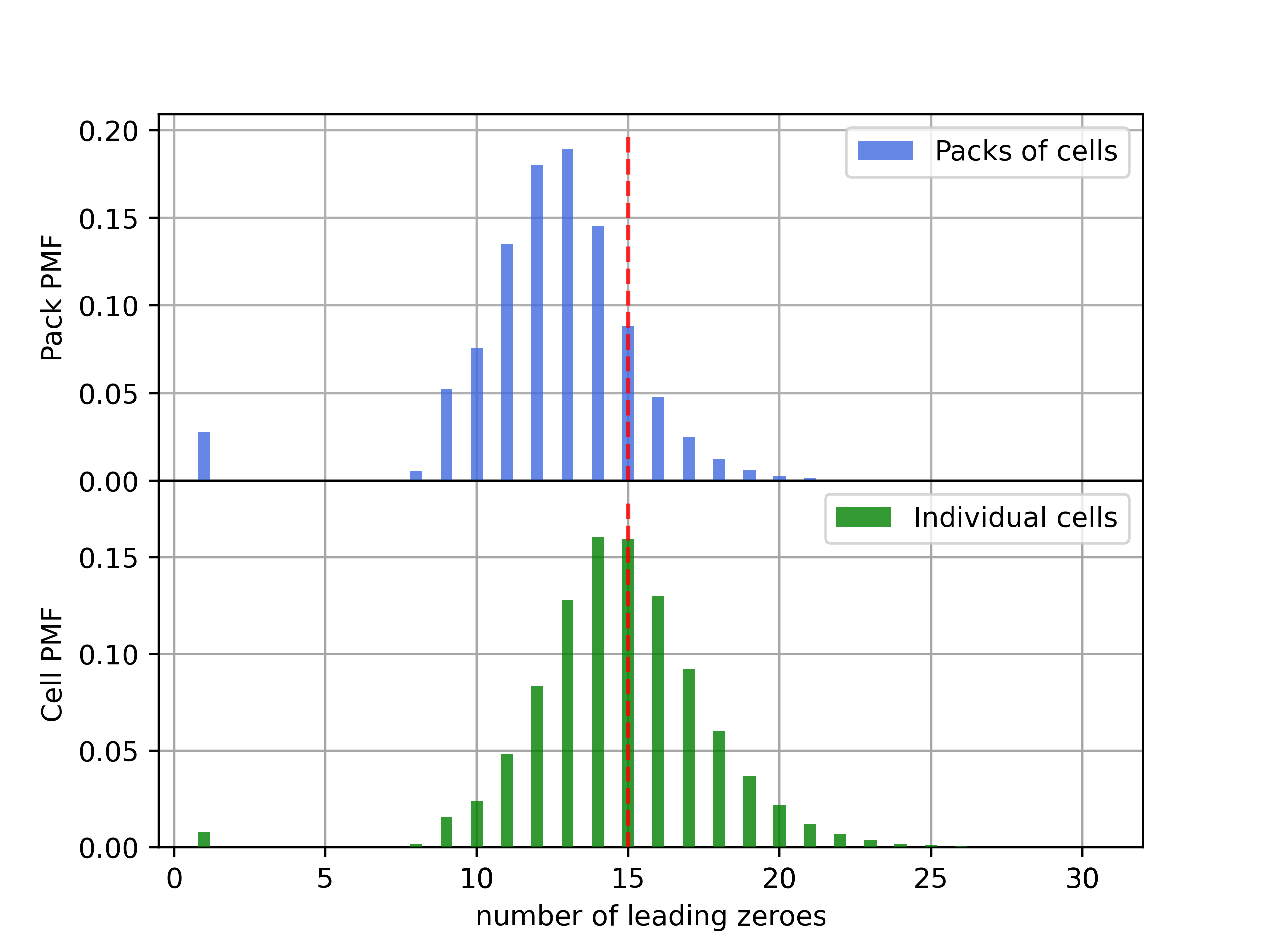}
	\caption{Histograms of number of removable leading zeroes (nLZ) on pack basis (top) and cell basis (bottom) on the double precision density field of Extreme-Horizon. In red dashed line is the threshold of 4 bits for the PCP4 encoding.}
	\label{fig:exh280_histo_rho}
\end{figure}

\firstReview{At the encoding stage, we need to encode the residues of the prediction errors of all children cell values as well as the number of removed leading zeroes. On figure \ref{fig:exh280_histo_rho} we show the probability mass function (PMF) of the number of leading zeroes on cell basis and on pack basis on the double precision density scalar field of the Extreme-Horizon simulation (12800 density datasets). Since residue bits are close to being uniformly random, we cannot compress them well. If we compute the compression ratio in this example, without adding the encoding of the number of removed leading zeroes (nLZ), we obtain 1.303 on cell-by-cell basis and 1.237 on pack basis which is expected since more bits can be remove in the first scenario. Nevertheless, the number of removed leading zeroes for each cell or each pack needs to be encoded for the decompression stage. The shannon's entropy of the nLZ field on a cell-by-cell basis is 3.5 and 3.2 on a pack basis, this smaller value being also an expected result since we smooth the nLZ value in each pack. Nevertheless, the total number of nLZ values to encode in the cell-by-cell case is, in 3D, 8 times the total number of nLZ values to encode in the pack case. Therefore, even by adding another stage of compression of the nLZ values, we cannot hope to achieve a better compression ratio even if more bits can be removed from the data. If we take into account the encoding of both the nLZ values and the residues, the compression ratio drops to 1.217 for the the cell-by-cell case, while it only drops to 1.225 for the pack case. What is more, the cell-by-cell approach would require 8 times more computations of leading zeroes and an additional compression step to barely achieve a similar compression ratio. Thus, we chose to compute and encode the nLZ value once per pack of children as shown at the encoding stage on figure \ref{fig:fsp_encoding}.\\

After computing the PMF of scalar field data, we notice that setting the number of removed leading zeroes on 4 bits by default is a good balance between compression speed and compression ratio. In the example of the Extreme-Horizon density field, increasing to 5 bits will only increase the compression ratio by 0.23 \% but reduce the compression speed by almost 8\%. Indeed, the nLZ computation is a key point of the algorithm performance and it requires more if-branches to compute a higher value. Finally, after the 4 bits used to encode the nLZ value, all residues $s^{'}_{i}$ are encoded, see figure \ref{fig:fsp_encoding}.}
 
According to the number of bits used for the number of removed leading zeroes, the theoretical maximum compression ratio in dimension $d$ is given by ($E_{enc}$ is the number of bits, 64 for double precision and 32 for single precision):
\begin{equation}
	C^{d}_{max} = \frac{E_{enc} \cdot 2^{d}}{(E_{enc}-15) \cdot 2^{d} + 4}
\end{equation}

Therefore, the maximum compression ratio achievable for single precision data is 1.829 and 1.293 for double precision with the default encoding. \firstReview{Nevertheless, we could expect this method to yield poor performance in the case of a difference of sign between a parent cell value and one of its child cells.} Indeed, the very first bit of a negative value will be set to one thus no leading zeroes will be removed. This could be the case for example with a small velocity field component varying around zero. Finally, our delta compression algorithm works for intensive variables but could be extended to extensive variables if the predicted values for the child cells were taken as the value of the parent cell divided by the numbers of child cells. If the number of child cells was a multiple of 2, the mantissa of the prediction error would not be modified (only the exponent part would be bitwise right-shifted), which could still lead to a lossless compression algorithm. The RAMSES code using only intensive variable AMR scalar fields, this extension of the Parent-Child predictor function is beyond the scope of this work.\\

\firstReview{For the float data decompression stage, the AMR grid refinement array is used level-by-level to guide the delta decompression algorithm through the grid from top to bottom since the scalar fields values at level $l$ are necessary as predictor values to decompress scalar field values at level $l+1$.}

\section{Results and discussion}
\label{sec:results}
In this section we present the results \firstReview{ of the lightAMR format conversion} as well as tree pruning and compression results obtained on different real simulation \firstReview{snapshots}: FRIG\footnote{\href{http://www.galactica-simulations.eu/db/STAR_FORM/FRIG/\#Simu_FRIG6_ZOOM7}{http://www.galactica-simulations.eu/db/STAR\_FORM/FRIG}}, snapshot at 10.05 Myr and 1675 coarse steps, \citep{Hennebelle2018} composed of 4096 domains, ORION\footnote{\href{http://www.galactica-simulations.eu/db/STAR_FORM/ORION/}{http://www.galactica-simulations.eu/db/STAR\_FORM/ORION}} snapshot at 1.26 Myr and 2125 coarse steps, \citep{Ntormousi2019} composed of 512 domains and Extreme-Horizon\footnote{\href{http://www.galactica-simulations.eu/db/COSMOLOGY/EXTREME_HORIZONS}{https://www.galactica-simulations.eu/db/COSMOLOGY/EXTREME\_HORIZONS}}, snapshot at $Z\sim1$ with 11866 coarse steps, \citep{Chabanier2020}, composed of 12800 domains. \firstReview{ Since, post-processing tools handle RAMSES data on a per-domain basis and that lightAMR is a self-consistent data model also on a per-domain basis, we consider a dataset as the data of one domain}. For the tree pruning algorithm we show the minimum, maximum and average value for different metrics that are computed on a per-domain basis. 




\subsection{Tree pruning results}
\label{subsec:results_pruning}

The table \ref{tab:pruning_rate_results} gives the results of the tree pruning applied on the different datasets. The algorithm is 2.3 to 3.4 times more efficient on FRIG and ORION data than on Extreme-Horizon. FRIG and ORION runs are zoom simulations \firstReview{(nested grids with increasing spatial resolution)} where most of the fine resolution grids can be found in the center or mid-plane of the simulation box, whereas Extreme-Horizon is a cosmological simulation where finest AMR levels are more homogeneously distributed in the  box. These results are consistent with the fact that RAMSES is more memory efficient in cosmological runs than in zoom simulations. Nevertheless, the tree pruning gain is applied on the whole simulation data therefore, 11.42\% on Extreme-Horizon means that the size is reduced by about 240 Gb. Moreover, tree pruning is done in a memory friendly way within a negligible time (\firstReview{a few microseconds on the data of an MPI process}) . Therefore this step has a significant impact on data volume at almost no cost. Actually, the time spent doing tree pruning is correlated with the number of coarse cells and the position of those coarse cells in the array.\\

\firstReview{The efficiency of the tree pruning algorithm is far from intuitive in most cases (at least with RAMSES AMR grids). It depends on the maximum level of refinement, on the number of domains, on the geometrical convexity of the subdomains, on their surface to volume ratio, and is greatly impacted by the 2:1 balance constraint in RAMSES required by its internal (hydro) numerical schemes. The tree pruning rates can span a wide range within a factor 3.25, even for different domains of the same simulation snapshot (see table \ref{tab:pruning_rate_results}).}

\begin{table}[!ht]
	\centering
	\renewcommand{\arraystretch}{1.1}
 	\begin{tabular}{l|p{0.07\textwidth}p{0.07\textwidth}p{0.07\textwidth}p{0.05\textwidth}}
  	\hline
	 Simulation & Min & Max & Global \\
  	\hline
  	FRIG  & 20.05 \% & 65.42 \% & 38.65 \% \\
	ORION & 17.23 \% & 47.32 \% & 26.35 \% \\
	ExH   & ~~7.00 \% & 23.24 \% & 11.42 \% \\
  	\hline
 	\end{tabular}
 	\caption{Fraction of removed cells on average by the tree pruning algorithm on Frig, Orion and Extreme-Horizon {\color{red}collections of datasets}. Minimum and maximum values are on domain (dataset) basis and global value is for the entire simulation snapshot.}
  	\label{tab:pruning_rate_results}
\end{table}

%

\subsection{LightAMR grid boolean arrays compression results}
\label{subsec:results_amr_comp}

We compared our compression speeds and ratios to commonly used and open source libraries : Zlib \footnote{\href{https://zlib.net/}{Zlib website}}, LZ4 \footnote{\href{https://lz4.github.io/lz4/}{LZ4 website}} and Snappy\footnote{\href{https://google.github.io/snappy/}{Snappy}}. The compression ratio is simply the ratio between the uncompressed and the compressed buffer sizes. We run this compression benchmark on the Extreme-Horizon datasets and compare the compression speed and ratio of the cumulated grid refinement and ownership mask arrays after tree pruning, see table \ref{tab:comp_benchs}. Domain datasets are processed sequentially, the uncompressed and compressed sizes are accumulated as well as the times to compress them. The compression speeds and ratios are then computed and reported in table \ref{tab:comp_benchs}. While our single threaded algorithm is slower than the LZ4 package, we are able to achieve a higher compression ratio. The zlib at level 9 (best compression ratio) achieves a much higher compression ratio with the grid refinement array but at the cost of a significantly reduced speed. The compression speed of all those algorithms are linked to the layout of the uncompressed array, i.e. consecutive pack of zeroes or ones. With the RAMSES current parallelism (one MPI process per core) we barely have to compress arrays greater than a few millions of cells, even in the Extreme-Horizon use case. Therefore, a compression speed of 374 Mb/s is enough to compress the entire lightAMR arrays of the most populated MPI process of the Extreme-Horizon dataset in less than 8 ms (domain 427 with 3059513 cells after tree pruning).\\

\begin{table}[!h]
	\centering
	\renewcommand{\arraystretch}{1.1}
 	\begin{tabular}{l|p{0.07\textwidth}p{0.07\textwidth}|p{0.07\textwidth}p{0.07\textwidth}}
  	\hline 
  	~ & \multicolumn{2}{c}{grid refinement} & \multicolumn{2}{c}{Ownership mask} \\
	 library & Ratio & Speed (MB/s) & Ratio & Speed (MB/s) \\
  	\hline
  	CPS52       & 23.95          &  ~~373.58        & \textbf{11546.10} &  ~~835.47 \\
  	LZ4         & 17.94          & \textbf{1255.15} &  ~~~~249.49       & \textbf{6864.82} \\
  	Zlib 1      & 26.00          & ~~233.93         &  ~~~~225.47       &  ~~289.31 \\
  	Zlib 9      & \textbf{48.13} & ~~~~~~7.46       &  ~~~~973.34       &  ~~155.22 \\
  	Snappy      & 12.00          & ~~918.88         &  ~~~~~21.29       &  1350.23\\
	\hline
 	\end{tabular}
 	\caption{Benchmark of compression of the lightAMR grid refinement and ownership mask arrays of the entire Extreme-Horizon snapshot using different libraries CPS52 (ours), the zlib version 1.2, the LZ4 package version 1.9.3 and Snappy 1.1.7. Benchmark run on a machine with a Intel Xeon Gold 5118 CPU @ 2.30 GHz.}
  	\label{tab:comp_benchs}
\end{table}

%
\begin{table}[!h]
	\centering
	\renewcommand{\arraystretch}{1.1}
 	\begin{tabular}{l|p{0.07\textwidth}p{0.07\textwidth}|p{0.07\textwidth}p{0.07\textwidth}}
  	\hline 
  	~ & \multicolumn{2}{c}{grid refinement} & \multicolumn{2}{c}{Ownership mask} \\
	 Simulation &  Ratio & Speed (MB/s) & Ratio & Speed (MB/s) \\
  	\hline
  	FRIG  & 39.36 & 1080.88  & ~~1627.19 & 1447.18 \\
	ORION & 48.06 & 1216.33  & ~~4397.45 & 1750.51 \\
	ExH   & 23.95 & ~~375.58 & 11546.10  & ~~835.47 \\
	\hline
 	\end{tabular}
 	\caption{Compression ratios and speeds of lightAMR boolean arrays for FRIG, ORION and Extreme-Horizon collections of datasets using CPS52 algorithm. Benchmark run on a machine with a Intel Xeon Gold 5118 CPU @ 2.30 GHz.}
  	\label{tab:comp_desc_results}
\end{table}

%

The compression ratios and speeds are much higher with the ownership mask array because it contains very large packs of successive zeroes (cells that belong to the domain). This is due to the fact that boundaries generally happen on the border of a domain but also because the tree pruning algorithm removes ghost nodes. In addition, our algorithm performs much better at compressing such an array than the other tested libraries because of encoding large packs in base 52 produce a very small pattern.

%

%

One can notice that our encoding values are stored on 1 byte even if there are only ranging from 1 to 63, and therefore would require only 6 bits thus a potential gain of 25\% in compressed size. In addition to that, computing the average Shannon entropy on 1 byte 'word' of all the AMR description array on an entire simulation like Extreme-Horizon leads to a value of 4. Which means that on average only 4 bits could be used to encode the information contained in one 'word'. The reason is simple, this is because all the number from 1 to 63 are not necessaryly used for the encoding. Actually, by computing exactly the number of bits that would be required to strictly encode bit to bit the information leads to a potential gain on average of 45\% on Extreme-Horizon and thus to reduce the size on disk of the AMR data by 45\% to less than 420 MB. We did not implemented such an optimization because it adds a step that will increase the time for compression for an AMR structure that is no longer significant compared to the size of the physical fields. And it will increase the complexity to rapidly retrieve AMR level seeds but also the direct access of cells states in the compressed data. Finally, in the case of Extreme-Horizon datasets, the AMR part of the data is now only representing 0.05\% of the total simulation snapshot volume. Table \ref{tab:comp_desc_results} compare the compressed lightAMR format to the legacy RAMSES AMR format.\\

One important remark is that most data arrays stored in a RAMSES checkpoint are no longer stored in this new lightAMR format, without any loss of information on the AMR grid structure itself. \firstReview{The large compression ratios presented in table \ref{tab:comp_desc_results} are mainly due to the switch to a boolean array description and the omission of this redundant meta-information, (e.\ g. neighbour cell indices, cell centers, parent cell indices) which can be easily and efficiently reconstructed on-the-fly by post-processing tools. The tree pruning step and the CPS52 encoding further increase this AMR file size compression ratio although in a less significant way.}

\begin{table}[!h]
	\centering
	\renewcommand{\arraystretch}{1.1}
 	\begin{tabular}{l|p{0.10\textwidth}p{0.12\textwidth}}
  	\hline
  	 Simulation & \centering Ratio & File size (MB) \\
  	\hline
  	FRIG  & \centering 1485.0 & ~~~~~~20 \\
	ORION & \centering 1500.0 & ~~~~~~~~6 \\
	ExH   & \centering ~~583.3 & ~~~~840 \\
    \hline
 	\end{tabular}
 	\caption{Chained ratio on RAMSES collections of datasets of tree pruning and CPS52 (left) and resulting lightAMR file size for the AMR grid description (right).}
  	\label{tab:comp_desc_results}
\end{table}

%

\subsection{LightAMR scalar field float data compression results}
\label{subsec:results_float_comp}

Most of the data volume of an AMR simulation dataset is due to scalar fields represented as floating point data. The lossless compression algorithm takes the grid refinement array and an associated scalar field float data array as input, both after being processed by the tree pruning algorithm. We used a maximum number of removed leading zeroes of 15 (4 bits encoding) and the compression is single threaded. First, we compare in table \ref{tab:bench_comp_float} the compression performances of the different libraries on \firstReview{the whole density field (double precison, 12800 datasets) from the Extreme-Horizon datasets}. We also integrate in the benchmark the Zfp library \cite{zfp}, which is  more specific to floating point data compression. We show that our algorithm achieve a higher compression ratio on this dataset due to the use of the topology information from the AMR mesh. It was expected that dictionary compression approach used in LZ4, Snappy and Zlib cannot efficiently compress floating point data. Nevertheless, since the Zlib also use an entropy encoded, it is able to achieve a few percent of compression but with very slow speed compared to our algorithm. \\


\begin{table}[!h]
	\centering
	\renewcommand{\arraystretch}{1.1}
 	\begin{tabular}{l|p{0.07\textwidth}p{0.12\textwidth}}
  	\hline 
	 library & Ratio & Speed (MB/s) \\
  	\hline
  	PCP4        & \textbf{1.230} &         1085.58  \\
  	LZ4         &         0.996  & \textbf{2935.65} \\
  	Zlib 1      &         1.060  &       ~~~~25.24  \\
  	Zlib 9      &         1.070  &       ~~~~21.17  \\
  	Snappy      &         1.000  &        ~~990.65  \\
  	Zfp         &         1.097  &       ~~~~72.03  \\
  	\firstReview{Zstandard}   &  1.061  &       ~~585.85 \\
	\hline
 	\end{tabular}
 	\caption{Benchmark of compression of the entire density field (12800 datasets) expressed according to the lightAMR format of Extreme-Horizon snapshot using different libraries: PCP4 (ours), the zlib version 1.2, the LZ4 package version 1.9.3, Snappy 1.1.7., zfp 0.5.5 \firstReview{and Zstandard 1.5.1. \citep{Zstd}}. Benchmark run on a machine with a Intel Xeon Gold 5118 CPU @ 2.30 GHz.}
  	\label{tab:bench_comp_float}
\end{table}

On Extreme-Horizon, a single scalar quantity over one snapshot represents about $\simeq 160$ Gb (after tree pruning) thus a compression of 1.23 on the density field means a gain of about 30 Gb on disk. In table \ref{tab:data_compression_results} we show the result of the PCP4 algorithm on the density, x-axis velocity, and thermal pressure for the different RAMSES datasets. We achieve a good compression ratio on the three tested scalar fields of about 1.19 in the worst case scenario and 1.23 on the best case. The compression speed is quite regular and above 1 GB/s, we noticed that this compression speed is linked with the CPU frequency thus with a higher CPU frequency we generally achieve a higher compression speed. For example, on an AMD Ryzen 7 3700X @ 3.6 Ghz we were able to reach compression speed above 2.5 GB/s up to 3 GB/s on FRIG and ORION datasets.
\firstReview{ We expected one current limitation of our algorithm for scalar fields with an average value of zero (e.\ g. velocity component). Indeed, if a parent value is positive and one of the children value is negative, then the very first bit of the integer representation will be set to 1 (sign bit) and therefore no delta compression could be achieved on that octant/quadrant. This limitation can be easily circumvented by shifting all values with a constant parameter. However, the results on table \ref{tab:data_compression_results} show that the compression ratios for the velocity component are always at least as good as those of the density and pressure fields, highlighting the fact that the ocurrence of opposite sign between predicted value (parent cell value) and child value (child cell value) is rather uncommon in typical astrophysical use cases.}\\

The main reason that explains our high compression speed is that we do not need to evaluate a mathematical function to predict each value but we merely use the parent value which is already available in memory. Moreover, this parent value is used as a predictor for all the child's cells (4 to 8 cells depending on the number of dimensions). Furthermore, based on the formula of the compression ratio, removing the bit of sign plus the entire exponent on the 64 bits integer representation of a double precision float for a pack of cells leads to a theoretical ratio of 1.219. We generally achieve a ratio higher than this threshold according to the results in table \ref{tab:data_compression_results}, which means that almost every child cell value lies within a factor two of the parent cell value, which demonstrates the interest of using our predictor function. Since the basic idea of AMR is to refine a cell to avoid high variation of the physical field between adjacent cells, it was expected that in most cases we achieve at least this ratio. Nevertheless in the density field of FRIG, this is not the case mainly because a lot of shocks (turbulent interstellar medium modeling) are present in the simulation, a large number of steep density discontinuities  can lead to lower compression ratio. 


\begin{table}[!h]
	\centering
	\renewcommand{\arraystretch}{1.1}
 	\begin{tabular}{l|c|p{0.07\textwidth}p{0.12\textwidth}}
  	\hline 
	 Simulation & Quantity & Ratio & Speed (MB/s) \\
  	\hline
	FRIG  & $\rho$ & 1.191 & 1274.94 \\
	FRIG  & $v_x$  & 1.223 & 1079.23 \\
	FRIG  & $P_t$  & 1.212 & 1100.95 \\
	\hline
	ORION & $\rho$ & 1.226 & 1236.57 \\
	ORION & $v_x$  & 1.219 & 1101.36  \\
	ORION & $P_t$  & 1.224 & 1240.4   \\
	\hline
	Ex-H   & $\rho$ & 1.225 & 1059.78 \\
	Ex-H   & $v_x$  & 1.258 & 1025.72 \\
	Ex-H   & $P_t$  & 1.169 & ~~960.54 \\
 	\end{tabular}
 	\caption{Compression ratios and speeds for \textbf{double precision} fields: density, velocity x-axis component, thermal pressure, of FRIG, ORION and Extreme-Horizon using the Parent-Child predictor algorithm using 4 bits encoding (15 removed zeros at maximum), maximum theoretical ratio is 1.293. Single-threaded on a Intel Xeon Gold 5118 CPU @ 2.30 GHz.}
  	\label{tab:data_compression_results}
\end{table}

Since the float compression data algorithm process the AMR tree from the top to the bottom, from coarse cells to leaves, we can not access directly the value of a cell at a level $l$ without first decompressing the tree from level 0 up to level $l$. Therefore, we give in table \ref{tab:decomp_speeds}, the decompression speed for double precision fields. The decompression speed we obtain is on a par with compression speeds while the decompression remains memory friendly.

\begin{table}[!h]
	\centering
	\renewcommand{\arraystretch}{1.1}
 	\begin{tabular}{l|c|p{0.12\textwidth}}
  	\hline 
	 Simulation & Quantity & Speed (MB/s) \\
  	\hline
	FRIG  & $\rho$ & ~~~~911.58 \\
	FRIG  & $v_x$  & ~~~~931.49 \\
	FRIG  & $P_t$  & ~~~~911.60 \\
	\hline
	ORION & $\rho$ & ~~~~927.62 \\
	ORION & $v_x$  & ~~~~921.84 \\
	ORION & $P_t$  & ~~~~928.15 \\
	\hline
	Ex-H   & $\rho$ & ~~~~760.18 \\
	Ex-H   & $v_x$  & ~~~~720.01 \\
	Ex-H   & $P_t$  & ~~~~709.69 \\
 	\end{tabular}
 	\caption{Decompression speeds for double precision fields: density, velocity x-component, thermal pressure, of FRIG, ORION and Extreme-Horizon (full level decompression). Single-threaded on a Intel Xeon Gold 5118 CPU @ 2.30 GHz.}
  	\label{tab:decomp_speeds}
\end{table}

Since in RAMSES one can downcast some physical quantities to single precision in order to reduce data volume by a factor 2, we give in table \ref{tab:comp_res_simple} the compression speeds and ratios for single precision data. If the entire exponent part (8 bits) of a single precision float (mapped on a 32 bits unsigned integer) is removed during the delta compression step, the theoretical compression ratio is 1.362, while it was only 1.219 for a double precision float. As shown previously with double precision float compression, we are still able to remove in most of the datasets the entire sign plus exponent part of the bit representation in the single precision float compression. Since the sign and exponent represent a bigger part of the 32 bits representation, it naturally leads to higher compression ratios. Therefore, down-casting double precision physical scalar fields within RAMSES can not only yield a reduction factor of 2 (due to down-casting) but also an even more efficient compression using our PCP algorithm. Consequently, down-casting from double to single precision can be of great interest in order to significantly reduce data volumes if double precision is not mandatory for the post-processing workflow.

\begin{table}[!h]
	\centering
	\renewcommand{\arraystretch}{1.1}
 	\begin{tabular}{l|c|p{0.07\textwidth}p{0.12\textwidth}}
  	\hline 
	 Simulation & Quantity & Ratio & Speed (MB/s) \\
  	\hline
	FRIG  & $\rho$ & 1.301 & 1259.76 \\
	FRIG  & $v_x$  & 1.393 & 1286.26 \\
	FRIG  & $P_t$  & 1.349 & 1299.11 \\
	\hline
	ORION & $\rho$ & 1.401 & 1292.74 \\
	ORION & $v_x$  & 1.380 & 1306.45  \\
	ORION & $P_t$  & 1.384 & 1331.21   \\
	\hline
	Ex-H   & $\rho$ & 1.394 & 1209.06 \\
	Ex-H   & $v_x$  & 1.537 & 1217.75 \\
	Ex-H   & $P_t$  & 1.345 & 1232.63 \\
 	\end{tabular}
 	\caption{Compression ratios and speeds for down-casted fields to \textbf{single precision}: density, velocity x-component, thermal pressure, of FRIG, ORION and Extreme-Horizon using the PCP4 algorithm (15 removed zeros at maximum), maximum theoretical ratio is 1.829. Single-threaded on a Intel Xeon Gold 5118 CPU @ 2.30 GHz.}
  	\label{tab:comp_res_simple}
\end{table}

~\\
\firstReview{The PCP4 algorithm is based on parent value as predictor value, this algorithm cannot be used if coarse node value are not available. Nevertheless, the basic idea of the lightAMR data model is to be a post-processing dedicated data model. AMR post-processing tools heavily use LOD approach which requires node value in order to be efficient. Therefore, if no node value is provided, then the whole dataset must be loaded to compute node values through up-sampling, such an operation will introduce very large overhead even for simple task such as low resolution images of the simulation box. The typical compression ratios that we obtain more than compensate the 12.5 \% overhead required to store node values (in 3D cell-based AMR octree).}

\section{Conclusion}

Adaptive Mesh Refinement suffers from a lack of standardized format to describe the grid structure and therefore AMR codes generally use their own format or eventually the unstructured grid to describe the mesh. Unfortunately, using the unstructured grid description is highly inefficient with AMR meshes and produce very large file even for relatively small meshes. In this paper, we presented an extremely compact format designed for AMR meshes, called \textit{lightAMR}, for which we developed a set of dedicated data reduction and compression algorithms. To minimize data redundancy in RAMSES AMR grids, we developed a tree pruning algorithm able to remove between 11\% and 38 \% of cells in various astrophysical datasets. The large discrepancy of the obtained values is mainly due to the diversity of adaptive refinement layouts in our selected dataset simulation boxes.
On top of that, we implemented a lossless and  memory friendly compression algorithm specific to the lightAMR format that compresses the grid refinement/ownership mask arrays using base 52 continuous pack size, called CPS52. \firstReview{By switching to a boolean array description of the AMR grid in the lightAMR format, by omitting unnecessary meta-information and by} chaining the tree pruning and the CPS52 algorithms on the tested RAMSES AMR datasets, we obtain AMR mesh file size compression ratios ranging from 583.3 to 1500.0, thus reducing the AMR grid description to a negligible part of the overall data volume.\\

In addition, a lossless, memory friendly and efficient floating point data compression algorithm, called PCP, designed for lightAMR physical scalar fields  achieve a high compression ratio and speed compared to commonly used and open source libraries by taking advantage of the topology of the AMR structure. For double precision, we achieve a compression speed ranging from 960 MB/s to 1275 MB/s and a ratio between 1.17 to 1.26, on RAMSES datasets, with a sequential version of the algorithm and a Intel Xeon Gold 5118 @ 2.30 GHz. While down-casting physical scalar fields to single precision naturally reduces the size by a factor 2, it also increases the compression ratios that range from 1.30 to 1.54 with similar compression speeds. \\

Finally, for the tested astrophysical datasets, the use of the lightAMR format leads to  overall data reduction percentages of 62.26 \% on FRIG, 49.64 \% on ORION and 37\% on Extreme-Horizon with no loss of information. \firstReview{If a RAMSES user is ready to sacrifice the precision of the scalar quantity fields and lets the lightAMR format downcast float data to single precision, then he can expect even more significant data volume reduction percentages (83.2\% on FRIG, 81.3 \% on ORION, 75.8 \% on Extreme-Horizon). In the context of the three astrophysical simulation projects tested in this work, all the while using the same disk storage capacity, far more ambitious data output policies could have been considered, with data output done at higher frequencies (e.g. x5 on ORION, x6 on FRIG, x4 on Extreme-Horizon).} The lightAMR format, as well as compression algorithms, are compatible by design with \textit{level-of-details} (LOD) approach which is a very important technique used by post-processing tools to improve performance for both visualization and data analysis. 


\begin{table}[!h]
	\centering
	\renewcommand{\arraystretch}{1.1}
 	\begin{tabular}{l|ccc}
  	\hline
	~ &  FRIG & ORION & Ex-H\\
  	\hline
  	\small{RAMSES Ncells ($10^6$)} & 1288.15 & 445.28 & 23660.29 \\
  	\small{lightAMR Ncells ($10^6$)} &  790.28 & 327.93 & 20958.90 \\
  	\hline
  	\small{RAMSES AMR (GB)}  & 29.7~~ & 8.70  & 490     \\
	\small{lightAMR (GB)}   & ~0.020  & 0.006 & ~~0.84  \\
	\hline
	\small{RAMSES HYDRO (GB)}   & 113.4  & 46.9  & 2337.42    \\
	\small{lightAMR HYDRO (GB)} & ~~54.0 & 28.0   & 1781~~~    \\
    \hline
    \small{RAMSES HYDRO SP (GB)}   & 56.7 & 23.45 &  1168.71  \\
    \small{lightAMR HYDRO SP (GB)} & 24.0 & 10.42 & ~~683.22 \\
 	\end{tabular}
 	\caption{Recap of data volume reduction and compression on RAMSES simulation datasets. LightAMR data takes into account the tree pruning algorithm. The first two lines gives the total number of cells before and after tree pruning. The AMR grid data volume using both the tree pruning and the CPS52 compression algorithms and the HYDRO data volume is reduced using the tree pruning and the PCP4 compression algorithms. The last two lines show the data volume reduction when float data are down-casted to  single precision (SP).}
  	\label{tab:recap_results}
\end{table}

\secondReview{Nevertheless, we must insist on the fact that the lighAMR data structure is designed to significantly reduce the data storage impact of RAMSES (or any other fully threaded octree code) simulations as presented in the paper and must be seen as a "storage data structure". For analysis purposes, a "computational-friendly" data strucuture, \citep{Bangerth2011}, should be constructed from the lightAMR data in order to achieve good performance. In our analysis workflow, for example, we use vectors or hashmaps of leave cells only, allowing us to achieve high performance data analysis with an hybrid multiprocessing / multithreading parallelism but this discussion is beyond the scope of this article.}

\section{Future work}
\secondReview{In the context of the standardisation process of the data format, the custom CPS52 run-length encoder will be updated in the near future to match more closely other mainstream run-length encoding schemes found in the literature, prior to the first release of LightAMR format.} While the detailed PCP4 compression is lossless, we will explore in future work lossy compression as a compromise to down-casting physical quantities to single precision in order to further reduce I/O volume. In order to further reduce the data volume for post-processing and since the lightAMR data model is self-consistent and self-describing even over one domain, we will explore the concatenation of multiple ligthAMR data from adjacent domains. \firstReview{Finally, we plan to use this updated version of RAMSES with the integrated lightAMR format and parallel I/O with Hercule library for a large and ambitious astrophysical simulation run on a pre-exascale architecture.}

\subsection*{Acknowledgements}
	The authors are thankful to P. Hennebelle and F. Bournaud for their useful comments on this manuscript. The authors acknowledge financial support from the European Research Council (ERC) via the ERC Synergy Grant {\em ECOGAL} (grant 855130).

%
   \bibliographystyle{elsarticle-num} 
   \bibliography{ms} 
%

%
%

\end{document}